\documentstyle[aas2pp4,flushrt]{article}

\newcommand{\gw}{GW46.4+5.5}
\newcommand{\mum}{$\mu$m}
\newcommand{\ihund}{$I_{100}$}

\newcommand{\isixty}{$I_{60}$}
\newcommand{\thund}{$\tau_{100}$}

\newcommand{\lb}{($l$, $b$)}
\newcommand{\degr}{$^\circ$}
\newcommand{\degrs}{$\!\!^\circ$}
\newcommand{\kms}{km~s$^{-1}$}
\newcommand{\vlsr}{$v_{\rm LSR}$}
\newcommand{\inhunit}{MJy~sr$^{-1}$~{(10$^{20}$~cm$^{-2}$)}$^{-1}$}
\newcommand{\cfunit}{cm$^{-2}$~{(K~\kms)}$^{-1}$}
\newcommand{\nh}{$N({\rm H})$}
\newcommand{\nhi}{$N({\rm HI})$}

\newcommand{\msol}{$M_\odot$}

\newcommand{\td}{$T_{\rm d}$}

\newcommand{\ta}{$T_{\rm A}^*$}
\newcommand{\tr}{$T_{\rm R}^*$}
\newcommand{\wco}{$W_{\rm CO}$}

\newcommand{\delvfwhm}{$\Delta v_{\rm FWHM}$}
\newcommand{\tbone}{$T_{\rm b,11}$}
\newcommand{\tbtwo}{$T_{\rm b,21}$}
\newcommand{\mcone}{KK99~3}
\newcommand{\mctwo}{KK99~4}

\begin{document}

 
\title{An Infrared and Radio Study of the Galactic Worm \gw}

\author{Kee-Tae Kim and Bon-Chul Koo}
\affil{Department of Astronomy, Seoul National University, Seoul
151-742, Korea}
 
 
\vskip -0.5cm
\begin{abstract}

In order to study the physical properties and origin of
the Galactic worm \gw, we have carried out
high-resolution ($\sim$3$'$) HI and CO (J=1$-$0) line observations
and analyzed available infrared and radio emission survey data.
\gw\ appears as a long ($\sim$8\degr),
filamentary structure extending vertically from the Galactic plane
in both median-filtered far-infrared and radio continuum maps.
The \isixty/\ihund\ ratio in \gw\ is estimated to be 0.29$\pm$0.05,
which is significantly higher than the value determined
for the solar neighborhood. The high ratio is consistent with
a hypothesis that the dust grains in the worms
have been processed by interstellar shocks.
The radio continuum emission from \gw\ has spectral index
$\alpha \simeq -$0.47 and does not correlate with \isixty\
except for emission at low ($|b| \leq$0.\degrs5) latitudes.
Thus, most of the radio continuum emission
is likely to be nonthermal.
Our one-dimensional HI observations show that the HI gas associated with
\gw\ is mainly at \vlsr$\simeq$15$-$40~\kms. The HI gas is clumpy and we
detected two molecular clouds associated with the HI peaks.
The molecular clouds have large internal velocity dispersions,
8.0 and 6.6~\kms, compared to their masses, 2.8$\times$10$^3$
and 1.7$\times$10$^3$~\msol, which implies that
they are not gravitationally bound.

Using the Leiden-Dwingeloo HI data,
we identify an expanding HI supershell associated with \gw,
which is centered on \lb$\simeq$(42\degr, 5\degr) with an
angular size of 14\degr$\times$22\degr\ 
(or 340$\times$540~pc$^2$ at 1.4~kpc).
The supershell appears between \vlsr$\simeq$18 and 40~\kms,
and decreases slowly in size as the velocity increases.
An averaged position-velocity diagram
reveals that
the supershell has a central velocity of $\sim$18~\kms, giving a
kinematic distance of 1.4~kpc, and
an expansion velocity of $\sim$15~\kms.
Assuming that it has been created by multiple stellar winds and
supernova explosions, we estimate its kinematic age and the energy
required to produce it to be about 5~Myr and
1.5$\times$10$^{52}$~ergs, respectively.
The structure is also visible in median-filtered radio
continuum maps, but not in the $ROSAT$ maps.
The observed molecular clouds might have condensed out of shock-compressed gas
in \gw\ because they are closely associated with the HI gas
in velocity as well as in position.
Their altitudes are 80 and 100~pc, respectively, higher than
the scale height of the thin molecular gas disk.
The physical properties of the clouds are very similar to those of the
high-altitude clouds observed recently in sensitive wide-latitude CO
surveys.
Our results suggest that
at least some of the high-altitude clouds might have
formed in Galactic worms (or swept-up HI shells and supershells).

\end{abstract}


\clearpage


\section{Introduction}

Galactic worms are filamentary, vertical structures which look
like worms crawling away from the Galactic plane in
HI maps (Heiles 1984). These structures
are found in the far-infrared (far-IR)
as well as in HI maps (Koo, Heiles, $\&$
Reach 1992 [KHR]). KHR identified
118 Galactic `worm-candidates' by cross-correlating
the 60, 100~\mum, and HI maps, and  statistically examined
their physical properties. The 118 structures were called worm
candidates because more than half of them are at high latitudes, i.e., 
they depart from the original morphological definition
that worms are vertical structures crawling out of the
plane regardless of their origins.
Individual features are wiggly and generally appear over a large
velocity interval. A noteworthy feature is that 35 among the 118
structures have corresponding structures
in 408~MHz continuum emission and 70\% of
the 35 structures appear to be associated with HII regions. The radio
continuum emission is likely to be thermal for bright worms that are
associated with HII regions.
Heiles (1984) first suggested that the worms might be vertical parts
of supershells which are quite extended in the z-direction or
which have broken through the thin ($H$~$\sim$~170~pc) gaseous disk
and are thus open at the top, i.e., ``chimneys'' (Norman $\&$ Ikeuchi
1989). Shells and supershells
have been found not only in the Galaxy but also in other
spiral and irregular galaxies in the Local Group
(Brinks $\&$ Bajaja 1986; Deul $\&$ den Hartog 1990; Kamphuis, Sancisi,
$\&$ van de Hulst 1991; Kim, S. et al. 1998). In some edge-on spirals,
worm-like vertical structures have also been directly observed
(Rand, Kulkarni, $\&$ Hester 1990, 1992; Swaters,
Sancisi, $\&$ van der Hulst 1997).
The most probable origin of the structures is understood to be
multiple stellar winds and supernova explosions in OB associations.
Such structures should be filled with tenuous, hot ionized gas.
The worms therefore could be pathways through which
both Lyman continuum photons and hot gas escape from the thin gaseous disk
so as to maintain
the diffuse warm ionized medium, which is widespread up to $\sim$1~kpc
from the Galactic midplane,
and the hot ionized gas in the halo (Reynolds 1991; Dove $\&$ Shull 1994;
Heiles, Reach, $\&$ Koo 1996).
Collisions of high-velocity clouds with
the galactic plane could also produce shells and supershells
in certain specific cases (Tenorio-Tagle 1981).

Notwithstanding their importance in understanding the global structure and
evolution of the interstellar medium, however, there are only a few
detailed studies on individual Galactic worms or chimneys
(M$\ddot{\rm u}$ller, Reif, $\&$ Reich 1987;
Normandeau, Taylor, $\&$ Dewdney 1996;
Maciejewski et al. 1996; Dennison, Topasna, $\&$ Simonetti 1997).
In this paper, we report the results of a multiwavelength study of
the Galactic worm \gw. \gw\ is a long,
filamentary structure stretching out vertically
from the Galactic plane. The morphology is similar to the North Polar
Spur (NPS) in the sense that the radio continuum structure is
systematically shifted by $\sim$0.\degrs5 from the HI (or far-IR)
structure (KHR).
It is likely that the NPS is composed of an inner radio continuum shell
surrounding an X-ray emitting region and an outer HI shell
(Borken $\&$ Iwan 1977; Heiles et al. 1980). Borken $\&$ Iwan (1977)
suggested that the NPS might be a very old ($\sim$10$^6$~years), large
supernova remnant (SNR), but that the X-ray emission
might be due to a reheating event, such as a recent ($\sim$10$^4$~years)
SN explosion.
KHR noted that the shift between the radio continuum and the HI emission
in \gw\ may be related to the origin of
the object as for the NPS. However, owing to  low ($\sim$30$'$) resolution, 
the data  were inadequate for the study of the correlation 
at different wavelengths.

We have carried out high-resolution ($\sim$3$'$) observations
of \gw\ in the HI 21~cm and CO J=1$-$0 lines,
and investigated the physical properties of HI and molecular clouds
associated with \gw.
Using available survey data in infrared and radio emission,
we show that \gw\ is the vertical wall of an expanding HI supershell.
The observations are described in Section 2 and
the results are presented in Sections 3, 4, and 5.
In Section 6, we examine the association between \gw\ and other large-scale
structures and consider the origin of \gw. We also discuss the origin of
the observed molecular clouds.
The last section contains our conclusions.

\section{Observations}

HI 21~cm line observations were undertaken
using the 305 m telescope at
Arecibo Observatory in 1990 October. The telescope has a half-power
beamwidth (HPBW) of 3.$'$3 and a main beam efficiency of 0.8 at 1.4 GHz.
We observed both circular polarizations simultaneously 
using two 1024 channel correlators with 5 MHz bandwidth each, 
and the velocity resolution was 2.03~\kms~after Hanning smoothing. 
Each spectrum was obtained by integrating for 1 minute
using frequency switching. We obtained a total of 9 one-dimensional cuts
through \gw\ at a set of constant Galactic latitudes 
given by 1\degr, 2\degr, $\cdots$, 9\degr. 
The spectra were taken at every 3$'$ in Galactic longitude.
The positions of our one-dimensional cuts are listed in Table 1.

CO J=1$-$0 line observations were made in 1994 February and 1995 March
using the 4~m telescope (HPBW~=~2.$'$5) at Nagoya University in Japan.
An SIS mixer receiver and a 1664 channel Acousto-Optical Spectrometer
(AOS)
with 40 MHz bandwidth were used. The velocity resolution was
0.66~\kms~after Gaussian smoothing. The system temperature
varied in the range 480$-$700~K during the observing sessions
depending on weather conditions and elevation of the source.
We first obtained 8 one-dimensional cuts in the same way as in the HI
observations.  Absolute-position switching
instead of frequency switching was used in order to avoid
contamination of the spectra
by the atmospheric CO emission. The reference positions were checked to
be free of appreciable (\tr\ $<$ 0.1~K) CO emission. The velocity was
centered at \vlsr~=~40~\kms. The on-source integration time
was 2 minutes and the typical rms noise level was
0.2~K per channel after smoothing.
The observed regions with the reference positions
are presented in Table 1.
We also mapped two regions bounded by
($l, b$)~=~(45.\degrs85$-$46.\degrs70, 2.\degrs60$-$3.\degrs55) and
(44.\degrs00$-$45.\degrs50, 3.\degrs40$-$4.\degrs45),
where CO line emission was detected in our previous observations.
We used absolute-position switching for the former and frequency switching
for the latter. 
We were able to use frequency switching for the latter cloud because,
when we observed the cloud in 1995 March, the velocity of the atmospheric CO
emission was very different from that of the cloud.
The beam separation was 3$'$. The central velocities were 27~\kms\ and 22~\kms,
respectively. The typical rms noise level was 0.2~K per channel after
smoothing.
The intensity scale was calibrated with respect to the standard
source S140 which was assumed to have \tr=20~K (Yang $\&$
Fukui 1992).

We also used the $IRAS$ coadded images (so called ``BIGMAP''),
which had been smoothed to 100~\mum\ resolution (HPBW$\simeq$4$'-$5$'$), 
and the Effelsberg 11 and 21 cm survey data (Reich et al. 1990; 
Reich, Reich, $\&$ F$\ddot{\rm u}$rst 1990).
The HPBW's of the Effelsberg 100~m telescope
are 4.$'$3 at 11~cm and 9.$'$4 at 21~cm, respectively.

\section{Infrared and Radio Continuum Properties}

\subsection{Infrared Properties}

Figure 1 shows the image of \gw\ in 100~\mum\ (grey-scale) and in 11~cm
continuum emission (contour). We removed a smooth background emission from the
far-IR and radio continuum maps by applying a median filter
of 6\degr$\times$2\degr. In far-IR emission,
\gw\ appears as an $\sim$8\degr\ long, filamentary
structure, stretching straight out from the Galactic plane.
It has a very narrow neck near b$\simeq$1.\degrs5 and broadens
to a width of about 2\degr\ above $b \simeq$2\degr.
Particularly interesting is a shell-like feature with a diameter
of $\sim$2\degr, centered on \lb$\simeq$(46.\degrs7, 5.\degrs3).
This structure, known as CTB~63, has been studied in the radio continuum by
Caswell, Landecker, $\&$ Roger (1989), who concluded that
it was not a discrete SNR because there was no sign of limb-brightening.
We can also see another hole centered near \lb=(45.\degrs8, 4.\degrs1).
The kinematics of the HI gas in this region will be discussed
in Section 4.

In order to study the physical properties of dust grains in
\gw, we examined the distributions of the 60/100~\mum\ color
temperature, \td, and 100~\mum\ optical depth, \thund.
Here we assume that the far-IR emission is
optically thin and the dust temperature is constant along
the line of sight, and use only data with \isixty$\ge$0.25 and
\ihund$\ge$1.00~MJy~sr$^{-1}$ to exclude
data points with peculiar values for \isixty/\ihund,
especially  on the border of \gw.
The derived \td~is dependent on $n$ which is
the index in the emissivity law,
$Q_{\rm abs}(\lambda) \sim \lambda^{-n}$.
For the graphite-silicate dust grain model of Mathis, Rumpl, $\&$
Nordsieck (1977), $n$ is known to be between 1 and 2 (Hildebrand 1983;
Draine $\&$ Lee 1984).
We adopt $n$=1.5 throughout the present paper.
In most regions of \gw,
\td~ranges from 26.1 to 28.7~K, namely
\isixty/\ihund=0.25$-$0.35. The average value of \isixty/\ihund\ for
\gw\ is 0.29$\pm$0.05, which is significantly higher than
0.21$\pm$0.02 for the solar neighborhood
(Boulanger $\&$ P\'{e}rault 1988) but is comparable to
the average value (0.28$\pm$0.03) of the Galactic worms (KHR).
For comparison, the NPS has infrared colors, including \isixty/\ihund\
and $I_{240}$/\ihund,
indistinguishable from those of the average interstellar medium
(Reach $\&$ Boulanger 1997).
Since about half the emission at 60~\mum\ is known to be attributed to
small, transiently heated grains (Draine $\&$ Anderson 1985;
D\'{e}sert, Boulanger, $\&$ Puget 1990),
the high value for \isixty/\ihund\ might suggest
that \gw\ contains a larger fraction of small grains. The enhanced
abundance of small grains appears to be consistent with a hypothesis
that the dust grains in the worms have been significantly processed
by interstellar shocks.

We derive the 100~\mum\ optical depth using
\thund= \ihund/$B_{\nu}$(\td), where $B_{\nu}$(T) is the Planck function.
The 100~\mum\ optical depth is still proportional to the amount of dust
along the line of sight (e.g., Kim, Lee, $\&$ Koo 1999),
although it does not represent the usual optical depth of
dust grains due to the contribution of small particles at 60~\mum.
In most regions of \gw, the range of
\thund\ is (1$-$10)$\times$10$^{-5}$, which
is an order of magnitude lower than that in dark clouds
(Wood, Myers, $\&$ Daugherty 1994).
There are two regions where the dust temperature is
considerably lower ($<$23~K) than the surrounding regions and where
the 100~\mum\ optical depth is very large ($\sim$4$\times$10$^{-4}$).
They are respectively located at
\lb$\simeq$(46.\degrs3, 3.\degrs0) and (45.\degrs1, 4.\degrs0).
These two regions occur where molecular gas is detected,
as we will show in Section 5.
We derive some far-IR parameters of \gw\ which are summarized in Table 2.
It should be noted that the measurements are dependent
on the size of the median filter used in the subtraction of a background
emission and the boundary values.

\subsection{Radio Continuum Properties}

The radio continuum structure extends outward
approximately vertically up to $b\sim$3\degr\ from the Galactic plane and
then overlaps with the far-IR structure at higher latitudes, so that the two
structures seem to be systematically shifted by 0.\degrs5$-$1.\degrs5 from
each other. The surface brightness of the radio continuum
declines dramatically near $b$=2\degr\, whereas it remains nearly constant
at higher latitudes.

The radio continuum emission from the interstellar medium
is divided into two distinct components:
thermal free-free and nonthermal synchrotron emission.
It is in general possible to distinguish the two components by
measurement of the spectral index $\alpha$ ($S_\nu \sim \nu^{+\alpha}$).
According to radio continuum observations of SNRs at various
wavelengths, the spectral index $\alpha$
ranges largely between $-$0.8 and $-$0.3
except for Crab-like SNRs, which have flatter spectra (Green 1991).
We estimate $\alpha$ for \gw\ using the 11 and 21~cm
continuum data from the formula

\begin{equation}
\alpha~=2~-~{\rm ln}(T_{\rm b,21}/T_{\rm b,11}) /
{\rm ln}(\lambda_{\rm 21}/\lambda_{\rm 11}),
\end{equation}

\noindent
where \tbone\ and \tbtwo\ are the 11 and 21~cm brightness
temperatures, respectively.
Since the 11 and 21~cm continuum data have different angular resolutions,
we convolved the 11~cm continuum map to
have the same angular resolution as the 21~cm map.
We plotted \tbtwo\ versus \tbone, and performed
a least-squares fit to the data. Here
the data points at $b \le 0.$\degrs5~are excluded because of a possible
contribution from HII regions in the Galactic plane.
The derived value for the spectral index
is $\alpha \simeq -$0.47, which is much steeper than
the value expected for thermal emission.

On the other hand, it was found that
there is a strong correlation between \tbone\ and
\isixty\ for HII regions
(Broadbent, Haslam, $\&$ Osborne 1989; Moon $\&$ Koo 1994).
The correlation has been used to decompose the 11~cm continuum into
thermal and nonthermal components in searching for SNRs.
We now examine whether or not there is a good correlation between
\tbone\ and \isixty\ for \gw.
Figure 2 is a plot of \tbone\ against \isixty.
Open circles and crosses represent data points below and above
$b$=0.\degrs5, respectively.
At $b \le 0.$\degrs5, we can see a fairly good correlation between
the two quantities, especially for data points with high
\tbone\ and \isixty. Additionally, the open circles lie near the solid
line, which is not a fit but represents the relation obtained by
Broadbent et al. (1989),
\tbone=(6.4$\pm$1.7)$\times$10$^{-3}$~\isixty,
where \tbone\ is in K and \isixty\ in MJy~sr$^{-1}$.
To the contrary, at $b > 0.$\degrs5,
there is virtually no correlation between
the two quantities. Almost all the data points, moreover,
are located to the left of the solid line, suggesting an excess of
11~cm continuum emission. It should be noted that \isixty\ is
the total intensity of the 60~\mum\ emission emanating from dust grains
associated with neutral as well as ionized gas,
and hence, if the continuum emission were thermal,
the data points are expected to fall to the right of the solid line.
Accordingly, the radio continuum emission from
\gw\ is likely to be nonthermal synchrotron emission.
However, we can not exclude a possibility that it is partially thermal,
because weak radio recombination line (RRL) emission was
detected at several positions in \gw\ (Heiles et al. 1996).
All the detected positions but one are situated at $b \le$1.\degrs5.
The central velocities of the RRLs are in the range
\vlsr=31.7$-$43.5~\kms, which is similar to that of the HI gas.

\section{HI Results}

Figure 3 displays the position-velocity diagrams of the HI 21~cm line
emission obtained in this work.
For simplicity, we show only the data between \vlsr$\simeq$$-$5 and 45~\kms.
In Section 6, we compare the far-IR maps with median-filtered HI
channel maps obtained from the Leiden-Dwingeloo survey data
(Hartmann $\&$ Burton 1997), and find that the HI gas between 
\vlsr$\simeq$18 and 40~\kms\
is associated with \gw. This velocity range is somewhat
wider than that determined by KHR using the Berkeley survey data,
\vlsr=25$-$38~\kms. By comparing Figure 3 with the far-IR
map in Figure 1, we see that the HI gas associated with \gw\ is
mainly at \vlsr$\simeq$20$-$40~\kms\ for $b \le$6\degr\, while it is mostly at
\vlsr$\simeq$10$-$30~\kms\ for $b \ge$7\degr.
As we describe in Section 6, however, the worm appears to be part of a
much larger supershell with a central velocity of 18~\kms. 
The corresponding kinematic distances are 1.4 and 11.4~kpc.
Here we use the Galactic rotation curve of Fich, Blitz, $\&$ Stark (1989),
which assumes $R_{\rm 0}$=8.5~kpc~and $\Theta_{\rm 0}$=220~\kms. 
We adopt the near
distance 1.4~kpc as the distance of \gw\ because the far distance
11.4~kpc yields the unacceptably large  worm-related parameters.

Figure 3 has a high ($\sim$1.3~pc) spatial resolution and shows that the
worm is clumpy. For example, at $b$=5\degr, there are two large
($\sim$50$'$ or $\sim$20~pc) `clumps' at ($l$, \vlsr)=(45.\degrs6, 32~\kms)
and (47.\degrs1, 29~\kms),
while at $b$=6\degr, we see several small clumps between $l$=46\degr\
and 48\degr. An interesting feature is a ring-like structure
which is centered on ($l$, $b$, \vlsr)$\simeq$(45.\degrs8, 4.\degrs0,
25~\kms).
This ring-like structure corresponds to the hole seen in the far-IR emission.
Thus, its morphology and velocity structure are very similar to
those of the HI shell either in expansion or contraction.
If the structure were an expanding HI shell, its expansion velocity
would be  $v_{\rm exp}$$\sim$5~\kms.
From the 100~\mum\ maps,
its size is estimated to be 37$\times$24~pc$^2$
(1.\degrs5$\times$1.\degrs0)
at the adopted distance of 1.4~kpc.

The HI column density, \nhi, is proportional to 
the integrated HI line intensity in the optically thin case.
In a separate paper (Kim et al. 1999),
we have found that the 100~\mum\ intensity
is tightly correlated with the HI column density in the region of \gw,
and that the estimated \ihund/\nh\ ratio is
\ihund/\nh$\simeq$1.3~\inhunit.
From the 100~\mum\ intensity map,
we determine the total mass of \gw\ to be 4.6$\times$10$^4$~\msol.
Our mass estimate includes a factor of 1.4 to account for
the cosmic abundance of He.

\section{Molecular Clouds Associated with \gw}

We have detected CO emission in five regions (Table 3).
They are marked by crosses in Figure 3.  Each molecular cloud has an
associated HI peak but its velocity is slightly different from that
of the HI peak. 
KK99~1 and KK99~2 are unlikely to be related to \gw\ because
their velocities, 60 and $-$10~\kms, are significantly different from the
velocity of the HI gas. \mcone\ and \mctwo\ are 
certainly associated with \gw. These regions correspond to the two
low-temperature regions mentioned in Section 3.
Their central velocities are, respectively, 27 and 22~\kms.
KK99~5 is also possibly related to the worm, although its central
velocity is somewhat lower than the velocity of the worm.

We have mapped the distributions of CO line intensity in
\mcone\ and \mctwo\, and Figure 4 shows the results.
Figures 4a and 4b have been obtained by integrating over the
velocity range \vlsr=18$-$35~\kms\ and 14$-$25~\kms, respectively.
The peaks are located at \lb=(46.\degrs35, 3.\degrs05) and
(45.\degrs10, 4.\degrs00). Peak emission spectra of the two molecular clouds
are shown in Figure 5. Their line widths (FWHM) are 7.9~\kms\ and
5.6~\kms, which are much greater than the line width of dense cores
of giant molecular clouds (GMCs), 1$-$3~\kms\ (Goldsmith 1987).
The large line widths seem
to be due to multiple components along the line of sight. The spectrum at
\lb=(45.\degrs10, 4.\degrs00) clearly shows double-peaked structure,
while the other does not show the apparent splitting of the line.
The line parameters at the strong cores of the two clouds are
listed in Table 4, in which
Galactic coordinates, central velocity,
line intensity (\tr), and line width (FWHM) are given.

\mcone\ is elongated along the
SE-NW direction and comprises two dense cores and an extended
envelope. There are two local peaks in the northwestern core.
The CO intensity in the dense core decreases steeply on the northern
side. The far-IR surface brightness also drops there.
According to the position-velocity maps,
the velocity dispersion increases abruptly
from the extended envelope to the cores,
but the central velocity remains approximately constant.

\mctwo\ consists of one dense core and
an extended envelope elongated in the SE-NW direction.
There is an interesting hole, centered on \lb$\simeq$(44.\degrs45,
4.\degrs10), with a diameter of $\sim$7$'$.
In the channel maps,
the central velocity appears to increase along the structure
from the northwestern part to the eastern end of \mctwo.
The core persists over
a large velocity interval of \vlsr=17.6$-$23.6~\kms.
Figure 6 is a position-velocity diagram along the dashed line 
in Figure 4(b), which shows clearly the velocity gradient.
However, we can also think of the possibility that
there are two components with a little velocity gradient within each
one, because the velocity dispersion of the western part is almost constant
and the velocity dispersion of the core region is about twice that of
the outer regions. 
In this case. the central velocities are
\vlsr$\simeq$19 and 22~\kms, respectively.
Each component is spatially distinct except for the core region, i.e.,
one component appears above $l$=44.\degrs10$-$45.\degrs30, whereas
the other is above $l$=44.\degrs90$-$45.\degrs60.
One plausible explanation for this velocity structure is a
cloud-cloud collision.

The integrated intensity of CO line emission, \wco, can be converted to the
molecular hydrogen column density, $N$(H$_2$), using

\begin{equation}
N({\rm H}_2)~=~X \cdot W_{\rm CO},
\end{equation}

\noindent
where $X$ is conversion factor in \cfunit.
By comparing HI, CO, and 100~\mum\ intensities,
we have derived 
$X$$\simeq$0.70$\times$10$^{20}$~\cfunit\ 
for \gw\ (Kim et al. 1999).
This value is much smaller than the estimated value for
molecular clouds in the Galactic plane, $X$=(1.8$-$4.8)$\times 10^{20}
$~\cfunit\ (Scoville $\&$ Sanders 1987), but comparable to
that for high-latitude clouds, $X$$\simeq$0.5$\times 10^{20}$~\cfunit\
(e.g., Weiland et al. 1986; de Vries et al. 1987;
Heithausen $\&$ Thaddeus 1990).
If we take the derived value, the masses of \mcone\ and \mctwo\ 
are 2.8$\times$10$^3$ and
1.7$\times$10$^3$~\msol, respectively.
We also estimate the virial masses of the clouds
using the formula

\begin{equation}
M_{\rm vir}~=~\frac{3 \beta \sigma_{\rm tot}^2}{\rm G} \overline{R},
\end{equation}

\noindent
where $\overline{R}$ is the geometric mean radius,
$\sigma_{\rm tot}$=\delvfwhm/{(8~ln2)}$^{1/2}$ is the
velocity dispersion, and  $\beta$ is a constant which depends on the shape
and density distribution of the cloud. Assuming a spherical cloud
with uniform density distribution ($\beta$=5/3), the virial masses are
1.1$\times$10$^5$ and 8.6$\times$10$^4$~\msol, respectively.
They are over an order of magnitude greater than the masses
calculated from \wco, so that
the molecular clouds do not appear to be gravitationally bound.
Table 5 lists the physical parameters of the molecular clouds.

\section{Discussion}

\subsection{Origin of \gw}

Worms are thought to be the vertical walls of supershells.
We have therefore attempted to find a large-scale structure associated with \gw\
at various wavelengths, including HI 21~cm line,
408, 820, 1420~MHz radio continuum, and X-ray emission.
In the Leiden-Dwingeloo HI channel maps, which are shown in Figure 7,
we could identify a shell-like structure associated with \gw.
The structure appears between \vlsr$\simeq$18 and 40~\kms.
Its diameter decreases slowly as the velocity increases.
Thus the structure seems to be the receding hemisphere of an expanding
HI supershell. The supershell is not
distinguishable at velocities lower than \vlsr=18~\kms.
We can think of two explanations for these observations:
either the supershell originally comprises only one hemisphere as 
for most other expanding supershells (Heiles 1979), or else
its approaching hemisphere is hidden by local HI gas.
On the other hand,
most parts of the supershell lie above the Galactic plane and its
center is located near \lb=(42\degr, 5\degr).
Its lower part,
which lies well below the plane at lower velocities,
moves upward as the velocity increases and then overlies the plane at
velocities higher than 33~\kms.
Hence, this expanding HI supershell might have been produced
by active events slightly above the Galactic plane and may have punched 
through the plane.
Its top part is coherent over the entire velocity range.
The supershell has an angular diameter of 14\degr$\times$22\degr\
at \vlsr$\simeq$18~\kms.
\gw\ is the eastern vertical wall of the supershell, while
GW39.7+5.7, another Galactic worm in the catalog of KHR, constitutes
the western wall. RRLs were also detected in GW39.7+5.7
and their central velocities range from
\vlsr=27.2 to 43.3~\kms, which are
in agreement with the velocity of \gw\ (Heiles et al. 1996).

Figure 8 exhibits a position-velocity map averaged over the Galactic
latitude range $b$=3\degr$-$7\degr. We can see a
hollow, centered on ($l$, \vlsr)$\sim$(42\degr, 18~\kms),
and two protuberances
near $l$=39\degr\ and 46\degr. The hole appears to be
surrounded by a dense HI shell,
which indicates that the HI gas in the hole might have been
swept into the shell.
The protuberances correspond to the two Galactic worms.
Using this position-velocity map, we estimate the central velocity
of the supershell to be 18~\kms, giving a kinematic distance of
1.4~kpc.
Thus its linear size would be 340$\times$540~pc$^2$.
The supershell seems to be situated on the near side of the Sagittarius
arm.
We also determine the expansion velocity to be $\sim$15~\kms.

The supershell is also visible
in median-filtered 408, 820, and 1420~MHz continuum maps (Haslam et al. 1982;
Berkhuijsen 1972; Reich $\&$ Reich 1986). The radio continuum
structure forms an almost complete shell and appears to be somewhat
smaller than the HI counterpart (Figure 9).
We derived the spectral index $\alpha$ of the structure using the 408
and 1420 MHz continuum data in the same
manner as in Section 3. For a region bounded by
\lb=(38\degr$-$50\degr, 1\degr$-$18\degr),
$\alpha \simeq -$0.86, implying that the radio continuum from
the supershell is nonthermal. This value is comparable to those
determined for large Galactic features, such as Loop I and Loop III
(Reich $\&$ Reich 1988), but the index is
much steeper than that obtained earlier by using the Effelsberg
11 and 21~cm continuum data.
This is probably because thermal emission is present at low
latitudes in \gw, as the detection of RRL emission suggests,
and the fraction of thermal emission to
nonthermal emission becomes greater at higher frequencies.
On the other hand, we could not find any obvious enhancement of X-ray emission
inside the supershell in the $ROSAT$ $\frac{1}{4}$~keV,
$\frac{3}{4}$~keV, and 1.5~keV bands (Snowden et al. 1997).
Assuming that all the HI emission in the velocity range 0$-$18~\kms\ 
arises between us and the supershell,
we estimate the foreground HI gas column density
toward the center of the supershell.
The derived values are (1$-$1.5)$\times$10$^{21}$ cm$^{-2}$,
which give optical depths
$\sim$0.5 in the $\frac{3}{4}$~keV band and $\sim$0.1 in the
1.5~keV band.
This suggests that the absence of X-ray emission in the bands is not
due to the absorption by the intervening interstellar gas.
Therefore, the supershell might be too old to have X-ray emission
harder than $\frac{3}{4}$~keV.
The $\frac{1}{4}$~keV emission from the supershell, if any,
would be absorbed completely by interstellar gas
because of its high ($\sim$10) opacity.

Figure 10 shows the distribution of molecular gas
between \vlsr=18 and 40~\kms,
obtained from the Columbia CO survey data (Dame et al. 1987).
We can see a diffuse, vertical filamentary structure extending at
$l$$\sim$46\degr\ and a gigantic molecular protuberance at
$l$$\sim$40\degr. The two structures are overlaid
on the lower walls of the HI supershell,
which  are depicted by light dotted lines.
There are several
optical HII regions (Fich et al. 1990), radio HII regions (Downes et
al. 1980; Lockman 1989), and massive star-forming regions
(Churchwell et al. 1990; Plume et al. 1992; Shepherd $\&$ Churchwell 1996)
inside and on the supershell.  These objects
reside mainly in the molecular protuberance.
Figure 10 displays their positions in the ($l$, $b$) and ($l$, \vlsr) planes.
Open squares, filled triangles, and crosses
represent optical,
radio HII regions, and massive star-forming regions, respectively.
Four SNRs, including G40.5$-$0.5, G42.8+0.6, G43.9+1.6, and
G45.7$-$0.4, are situated in this area (Green 1991). 
They are marked by ``$\times$'' symbols
in Figure 10. Here we have not included HII regions and SNRs,
such as W49A and W49B, 
that are known to be located at distances greater than 5~kpc 
based on HI absorption line observations.
We also find four open clusters, NGC~6709, NGC~6738, Berkeley~43,
and Berkeley~82 in the supershell.  Open circles denote them in Figure 10.
These clusters except NGC~6709 have hardly been studied. NGC~6709
is located near the center of the supershell. According to 
Hoag $\&$ Applequist (1965), this cluster contains no luminous star
earlier than B5 and its distance is about 1~kpc, which is roughly
comparable to the kinematic distance of the supershell.
However, NGC~6709 is unlikely to have served to produce the supershell because 
its age is known to be too large, $\sim$10$^8$~yr (Loktin $\&$ Matkin 1994),
in comparison with the kinematic age of the supershell derived
below. 
Therefore, the supershell might have been created
by multiple SN explosions occurring in
the eastern side of the gigantic molecular protuberance
or in the open clusters except NGC~6709 in its interior.
It seems possible, in addition, that the active events
have triggered some star-forming activity in the molecular cloud
complex as in the sequential star formation
model (Elmegreen $\&$ Lada 1977).

We estimate the energy required to produce the supershell and its age
using theoretical models.
Since the z-extent of our supershell is much
greater than the scale height of the HI gas disk,
we should consider the density gradient of the interstellar
medium perpendicular to the galactic plane.
With the use of numerical models several groups have investigated
the evolution of supershells, driven by stellar winds and
repeated SNe from an OB association, in various plane-stratified
gas distributions (Tomisaka $\&$ Ikeuchi 1986; Mac Low et al. 1989;
Igumentshchev, Shustov, $\&$ Tutukov 1990). The supershell in general
grows preferentially and so becomes extended in the direction perpendicular
to the galactic disk in this case.
With various energy input rates,
Igumentshchev et al. (1990) have modeled the evolution of supershells in
Gaussian and exponential gas disks.
The disks were assumed to have a scale height of 140~pc and a midplane
density of about 1~cm$^{-3}$.
They presented an empirical expansion law for the radii of supershells
in their equation (1):
$R_{\rm s} = 97 {(t/10^6~{\rm yr})}^{0.44}
{(\dot{E} / 2\times10^{38}~{\rm ergs~s}^{-1})}^{0.2}$~pc.
The observed radius and expansion velocity of the supershell then
suggest a kinematic age of
$\tau_k = 0.44 R_{\rm s}/v_{\rm ex} \simeq$5~Myr.
By substituting this age estimate into the expansion law,
we derive the energy input rate to be 1$\times$10$^{38}$~ergs~s$^{-1}$,
which means a total deposited energy of $\sim$1.5$\times$10$^{52}$~ergs.
The dynamical parameters of the supershell are summarized in Table 6.

\subsection{Origin of Molecular Clouds}

We found two molecular clouds associated with \gw.
The molecular clouds, \mcone\ and \mctwo, have similar
central velocities and internal velocity dispersions (Table 5).
As discussed above,
they are closely associated with the HI gas in velocity as well
as in position.
In the large-scale CO map at 0.\degrs5 resolution (Figure 10),
the molecular clouds appear as
a filamentary molecular structure extending out of the
Galactic disk.
The molecular filament is completely embedded in the HI filament.
Hence, it is likely to be a molecular counterpart of \gw, i.e., a
``molecular worm''.
This structure is similar to the one found in
the prominent worm GW23.0$-$1.6 by Dame (1996),
which extends up to a well-defined cloud
at $b$=$-$1.3\degr\ from an active star-forming region in the Galactic
plane.
However, Dame's molecular worm is twice as large as ($\sim$180~pc) and much
more massive ($\sim$6$\times$10$^4$~\msol) than ours.
Another two Galactic worms, GW30.5$-$2.5 and GW49.1$-$1.4, were found to
have associated molecular clouds at high altitudes (Heiles et al. 1996),
although their morphological associations were not investigated
in detail.

It has been suggested
that molecular clouds
might form in swept-up shells and supershells (cf. Elmegreen 1987 and
references therein). In this case,
individual clouds might form in local HI peaks
and on the whole the clouds would lie along the shell.
The distribution of local high-latitude clouds 
provides good examples (Gir, Blitz, $\&$ Magnani 1994).
McCray $\&$ Kafatos (1987) considered the formation of molecular clouds
inside a supershell produced in a uniform medium
by gravitational instability. According to their
results, for the case of the supershell being in the snowplow phase
as a result of radiative cooling or breaking through the galactic disk,
the gravitational instability begins at

\begin{equation}
t_1 \approx 12~E_{51}^{-1/15} n_{\rm o}^{-11/15}
a_{\rm s}^{4/5} R_{100}^{-7/15}~~~~{\rm Myr},
\end{equation}

\noindent
where $E_{51}$ = ($E_{\rm SN}$/10$^{51}$~ergs), $n_{\rm o}$ (cm$^{-3}$)
is the ambient density, $a_{\rm s}$ (\kms) is the magnetosonic speed,
and $R_{100}$=($R$/100~pc) is radius at which the transition from the adiabatic
phase to the snowplow phase occurs.
In the velocity range \vlsr$\geq$18~\kms,
an averaged value of the \nhi\ excess over the Galactic longitude range
of the supershell, compared to \nhi\ at other longitudes,
is $\sim$7$\times$10$^{20}$~cm$^{-2}$ (see Figure 8).
Assuming that this represents all the preexisting ambient gas swept up
within a line-of-sight radius of 170~pc, the original ambient gas
density would be $n_{\rm o} \sim$1.4~cm$^{-3}$.
If magnetic pressure can be neglected
$a_{\rm s} \sim$0.5~\kms\ and $R_{100} \sim$1,
$t_1 \sim$4.5~Myr.
So, it seems that \mcone\ and \mctwo\ could
have condensed out of shock-compressed gas in the HI supershell.

If we take $d$=1.4~kpc,
the altitudes of \mcone\ and \mctwo\ are 80 and 100~pc, respectively.
Since the scale height of thin molecular gas disk
is $\sim$70~pc (Bronfman et al. 1988, scaled to $R_{\rm 0}$=8.5~kpc),
they seem to be fairly far from the Galactic midplane.
Their physical properties are
intermediate between local high-latitude clouds and GMCs and so
are very similar to those of the high-altitude clouds,
which were identified recently by sensitive wide-latitude CO surveys
in the first Galactic quadrant (Dame $\&$ Thaddeus 1994; Malhotra 1994). 
The origin of the high-altitude clouds
is as yet poorly understood.
Our results suggest
the possibility that at least some
of the high-altitude clouds might have formed
in Galactic worms or swept-up HI shells and supershells.

\section{Conclusions}

It has been suggested that Galactic worms are the walls of 
supershells or chimneys. 
But their association has been rarely explored although there
have been some detailed studies on a few Galactic worms.
In this paper, we have shown that the well-defined Galactic worm \gw\ is
indeed the wall of a much larger ($\sim$340$\times$540~pc$^2$)
supershell. There is strong evidence suggesting that the supershell is
expanding at $\sim$15~\kms. The identification of the associated
supershell led us to derive an accurate kinematic distance (1.4~kpc) 
to \gw\ and its accurate physical parameters. 
Our results show that it is essential to
identify associated supershells for the study of Galactic worms and that
it is worthwhile to carry out similar analyses for other Galactic worms.
Our high-resolution HI and CO observations revealed the clumpy HI
structure of the worm and the associated molecular gas at HI peaks. The
physical properties of the molecular clouds are very similar to those of
the high-altitude clouds. The physical association of the molecular
clouds with \gw\ strongly suggests that they have formed within the
supershell. A more systematic study might reveal the general association
of high-altitude molecular clouds with Galactic worms (or HI
supershells).

\acknowledgements
 
We thank Yasuo Fukui, Akira Mizuno,
and the graduate students of the Department of
Physics and Astrophysics, Nagoya University for their help with
the CO line observations.
We are very grateful to Carl Heiles and Magdalen Normandeau
for several helpful comments and suggestions.
This work has been supported in part by 
Basic Science Research Institute Program, Ministry of Education, 
project 1998-015-000284.

\clearpage


\clearpage
 

\begin{deluxetable}{ccccc}
\tablewidth{0pt}
\tablecaption{POSITIONS OF ONE-DIMENSIONAL CUTS
AND OFF POSITIONS \label{tbl-1}}
\tablehead{
  &   & \multicolumn{2}{c}{$l$-Range}  &
\colhead{OFFs\tablenotemark{a}} \\
\cline{3-4}
 & \colhead{$b$} &
\colhead{HI 21 cm line} & \colhead{CO J=1$-$0 line} & \colhead{($l, b$)} \\ 
\colhead{Scan Number} & \colhead{(deg)} &
\colhead{(deg)} & \colhead{(deg)} & \colhead{(deg)}
}
\startdata
B1 & 1.0 & 42.5$-$48.0 & 43.0$-$45.5 & (45.0, 1.5), (43.5, 2.5) \nl
B2 & 2.0 & 43.0$-$45.7 & 43.0$-$46.0
& (45.0, 1.5), (43.5, 2.5) \nl
B3 & 3.0 & 42.5$-$48.0 & 44.0$-$47.0
& (43.5, 2.5), (45.8, 4.2) \nl
B4 & 4.0 & 43.0$-$47.5 & 44.0$-$46.5
& (47.5, 3.5), (45.8, 4.2) \nl
B5 & 5.0 & 42.5$-$50.0 & 45.0$-$48.0
& (46.5, 5.3), (49.0, 5.5) \nl
B6 & 6.0 & 42.5$-$51.5 & 46.0$-$48.5
& (46.5, 5.3), (49.0, 5.5) \nl
B7 & 7.0 & 42.5$-$51.5 & 46.5$-$47.5
& (45.5, 6.5) \nl
B8 & 8.0 & 42.5$-$51.0 & 46.5$-$47.5
& (48.0, 8.0) \nl
B9 & 9.0 & 42.5$-$51.5 & \nodata & \nodata \nl
\enddata
\tablenotetext{a}{OFF positions for CO J=1$-$0 line observations}
\end{deluxetable}

\begin{deluxetable}{lcl}
\tablewidth{0pt}
\tablecaption{FAR-IR PHYSICAL PROPERTIES OF \gw\ \label{tbl-2}}
\tablehead{
\colhead{Parameter} & \colhead{Value} & \colhead{Unit}
}
\startdata
Angular size in $l$ & $\sim$2       & degree\nl
Angular size in $b$ & $\sim$7       & degree\nl
Angular area        & $\sim$14      & square degree\nl
$<I_{60}>$          & 1.7$\pm$1.3 & MJy~sr$^{-1}$\nl
$<I_{100}>$         & 6.4$\pm$4.5 & MJy~sr$^{-1}$\nl
$<I_{60}/I_{100}>$  & 0.29$\pm$0.05 &  \nl
\td                 & 26$-$29       & K\nl
\thund              & (1$-$10)$\times$10$^{-5}$ &  \nl
\ihund/\nh\         & 1.32$\pm$0.02 & \inhunit\nl
\thund/\nh\         & 1.00$\pm$0.02 &
10$^{-5}$~{(10$^{20}$~cm$^{-2}$)}$^{-1}$\nl
$M_{\rm HI}$                & 4.6$\times$10$^4$ & \msol \nl
\enddata
\end{deluxetable}

\begin{deluxetable}{cccccc}
\tablewidth{0pt}
\tablecaption{REGIONS WITH CO J=1$-$0 LINE EMISSION \label{tbl-3}}
\tablehead{
\colhead{}   & \colhead{$b$} & \colhead{$l$-range} & 
\colhead{\vlsr\tablenotemark{a}} &
\colhead{\tr\tablenotemark{a}} &
\colhead{\delvfwhm\tablenotemark{a}} \nl
\colhead{Cloud} & \colhead{(deg)} & \colhead{(deg)} &
\colhead{(km~s$^{-1}$)} & \colhead{(K)} & \colhead{(km~s$^{-1}$)}
}
\startdata
KK99~1 & 1.0 & 43.50$-$43.70 & +60 & 2.5 & 2.1 \nl
KK99~2 & 1.0 & 44.50$-$44.75 & $-$10 & 2.1 & 2.9 \nl
\mcone\ & 3.0 & 45.30$-$46.65 & +27 & 3.0 & 7.8 \nl
\mctwo\ & 4.0 & 44.20$-$45.50 & +22 & 3.3 & 5.6 \nl
KK99~5 & 5.0 & 47.65$-$47.95 & +14 & 1.8 & 2.5 \nl
\enddata
\tablenotetext{a}{Line parameters at peak position}
\end{deluxetable}

\begin{deluxetable}{ccccc}
\tablewidth{0pt}
\tablecaption{CO J=1$-$0 LINE PARAMETERS AT PEAK POSITIONS \label{tbl-4}}
\tablehead{
   & \colhead{\lb} & \colhead{\vlsr} & \colhead{\tr} &
\colhead{\delvfwhm} \\
   & \colhead{(deg)} & \colhead{(\kms)} & \colhead{(K)} &
\colhead{(\kms)}
}
\startdata
\mcone\ & & & & \nl  \cline{1-1}
\mcone:P 1 & (46.35, 3.05) & 27.4 & 5.2 & 7.9 \nl
\mcone:P 2 & (46.20, 3.20) & 27.4 & 4.0 & 8.7 \nl
\mcone:P 3 & (46.10, 3.20) & 28.9 & 4.8 & 7.9 \nl
\mctwo\ & & & & \nl \cline{1-1}
\mctwo:P 1 & (45.10, 4.00) & 22.0 & 3.3 & 5.6 \nl
\mctwo:P 2 & (44.45, 4.20) & 18.5 & 2.8 & 2.5 \nl
\mctwo:P 3 & (44.20, 4.20) & 18.3 & 3.5 & 1.7 \nl
\enddata
\end{deluxetable}

\clearpage

\begin{deluxetable}{ccccccc}
\tablewidth{0pt}
\tablecaption{PARAMETERS OF THE MOLECULAR CLOUDS \label{tbl-5}}
\tablehead{
\colhead{} & \colhead{$<W_{\rm CO}>$} & \colhead{$\overline{R}$} & 
\colhead{\vlsr\tablenotemark{a}} & 
\colhead{\delvfwhm\tablenotemark{a}} & 
\colhead{$M_{W_{\rm CO}}$} & \colhead{$M_{\rm vir}$} \\ 
\colhead{Cloud} & \colhead{(K~\kms)} & \colhead{(pc)} & \colhead{(\kms)} & 
\colhead{(\kms)} & \colhead{(10$^3$ \msol)} & \colhead{(10$^4$ \msol)} 
}
\startdata
\mcone\ & 10.4 & 8.4 & 27 & 8.0 & 2.8 & 10.9 \nl 
\mctwo\ & ~4.6 & 9.7 & 22 & 6.6 & 1.7 & ~8.6 \nl
\enddata
\tablenotetext{a}{\vlsr\ is the center velocity and 
\delvfwhm\ is the FWHM of the composite spectral line}
\end{deluxetable}

\begin{deluxetable}{lcl}
\tablewidth{0pt}
\tablecaption{DYNAMICAL PROPERTIES OF THE HI SUPERSHELL \label{tbl-6}}
\tablehead{
\colhead{Parameter} & \colhead{Value} & \colhead{Unit}
}
\startdata
Size & 340 $\times$ 540       & pc$^2$\nl
Expansion velocity & $\sim$15 & \kms \nl
Kinetic age & 5 & Myr \nl
Ambient density & $\sim$1.4 & cm$^{-3}$ \nl
Total energy & 1.5 $\times$ 10$^{52}$ & ergs \nl
\enddata
\end{deluxetable}

\clearpage
\onecolumn
 

\begin{figure}
\figurenum{1}
\epsscale{1.0}
\vskip -3cm
\plotone{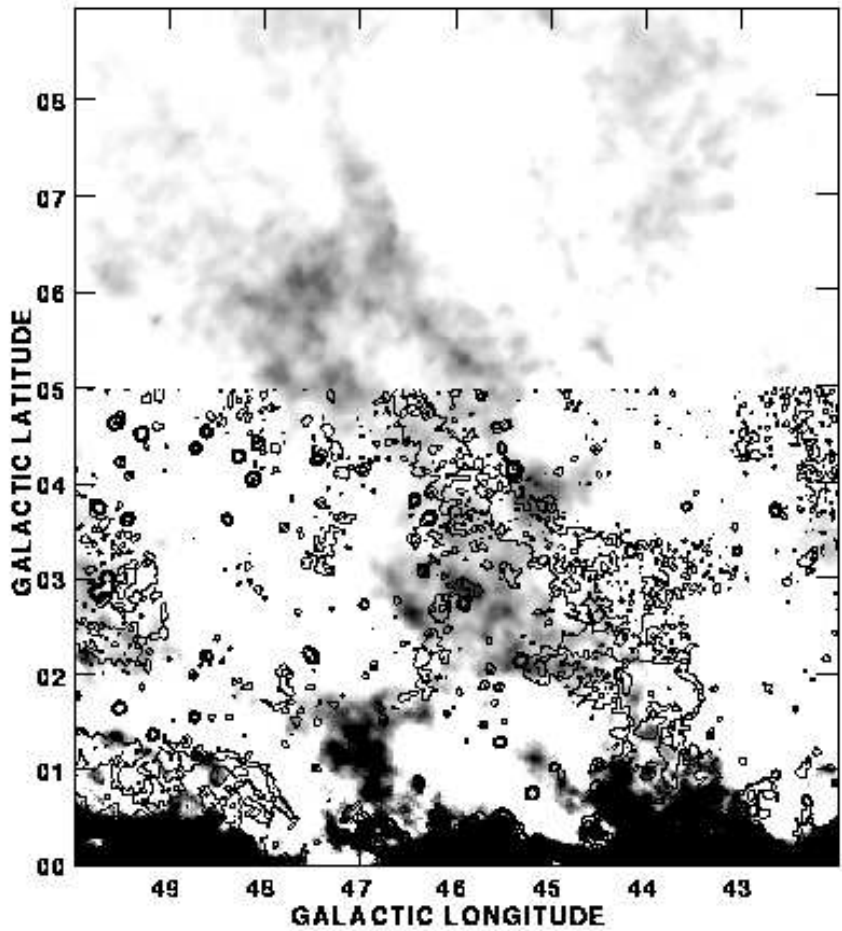}
\caption[figure1m.ps]{
100 \mum\ (grey scale) and 11~cm continuum (contour) images of \gw.
The far-IR structure extends straight up to $b$~=~7\degr\ from the
Galactic plane.
The radio continuum structure is shifted by about 1\degr\ from
the far-IR one between $b \simeq$1.\degrs5~ and 3.\degrs5. The grey
scale
intensity range is 1$-$30~MJy~sr$^{-1}$. Contour levels are 0.03, 0.1,
0.25, 1.0, and 3.0~K.}
\end{figure}
 
\clearpage
 
\begin{figure}
\vskip -13cm
\figurenum{2}
\epsscale{1.0}
\plotone{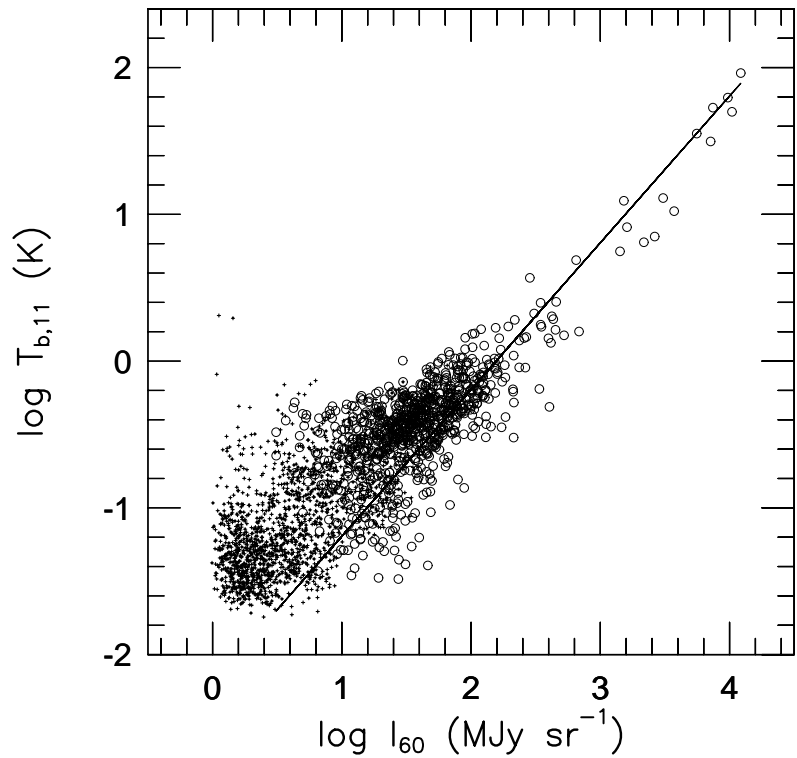}
\vskip -3cm
\caption[figure2.ps]{
The plot of 11~cm continuum brightness temperature,
$T_{\rm b,11}$,
against 60~\mum\ intensity, \isixty. Open circles and crosses
represent data points below and above $b$=0.\degrs5,
respectively. The solid line represents a
relation between the two parameters, derived by Broadbent et al. (1989),
for thermal emission.}
\end{figure}
 
\clearpage
 
\begin{figure}
\vskip -1cm
\figurenum{3}
\epsscale{0.8}
\plotone{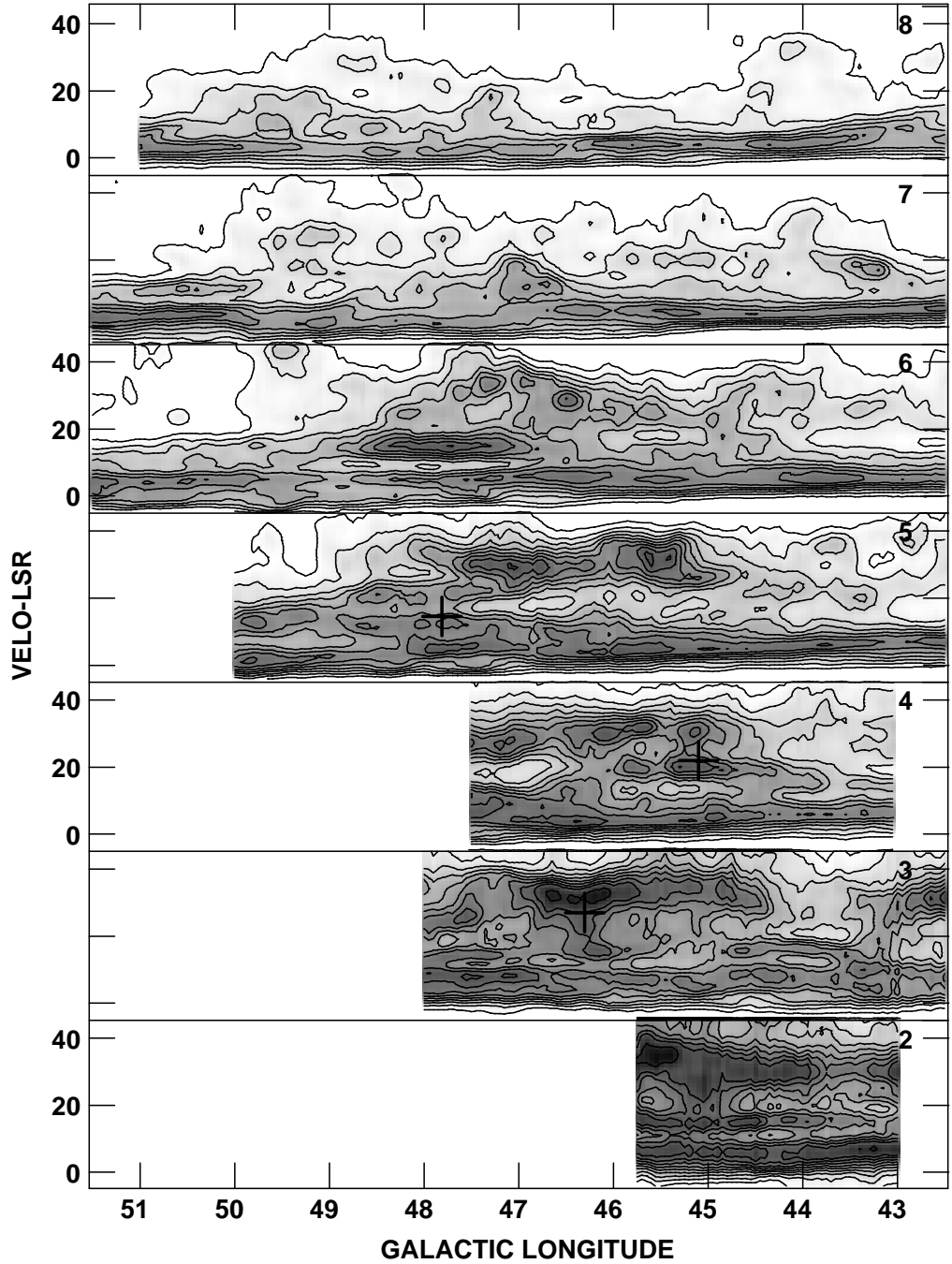}
\vskip -1cm
\caption[figure3.ps]{
Arecibo HI position-velocity diagrams at $b$=2\degr$-$8\degr.
The Galactic latitude
is indicated on the top right corner of each diagram. The HI gas
associated
with \gw\ is shown clearly above \vlsr$\simeq$20~\kms\
at $b \leq$6\degr.
The positions of molecular clouds detected in this work
are marked by crosses. The contours correspond to
5, 10, 15, 20, 25, 30, 35, 40, 45, 50, and 60~K.
The grey scale intensity range is 10$-$70~K.}
\end{figure}
 
\clearpage
 
\begin{figure}
\vskip -3cm
\figurenum{4}
\epsscale{0.8}
\plotone{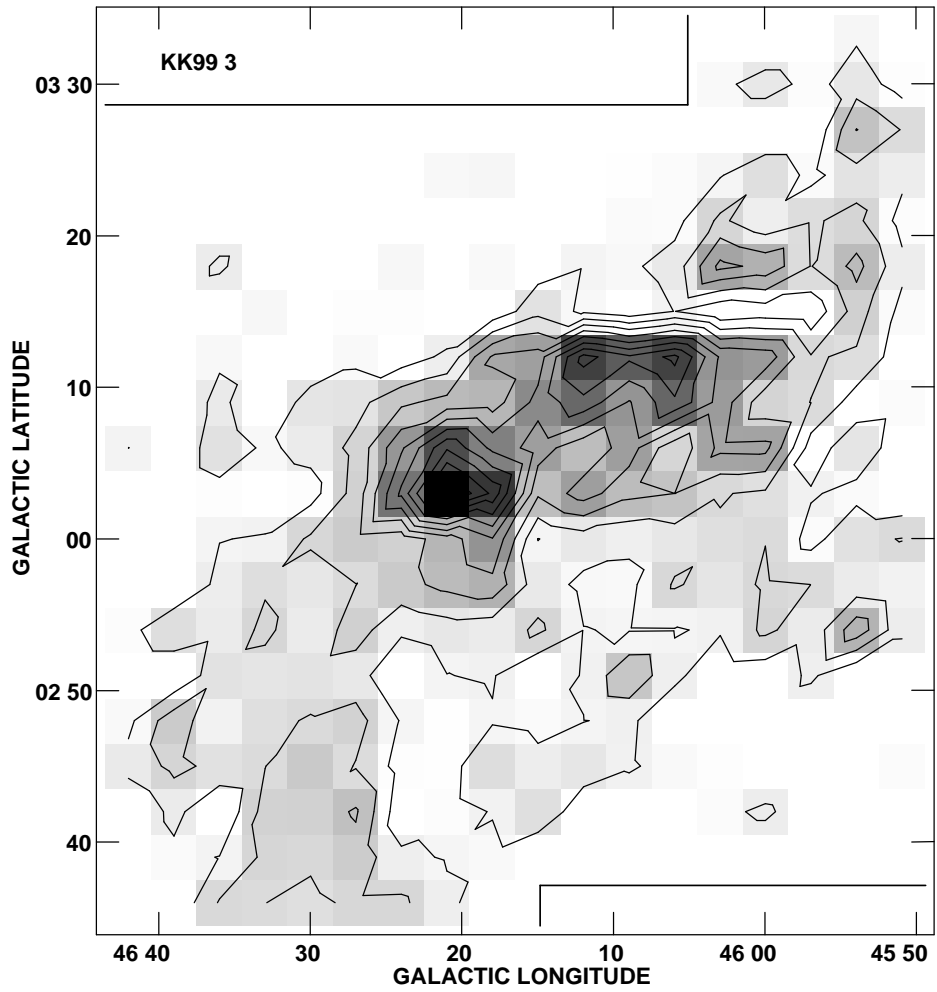}
\vskip -1cm
\caption[figure4a.ps]{
Integrated CO line intensity maps of molecular clouds 1 and 2.
(a) \mcone. The grey scale flux range is 2$-$44~K~\kms.
The distribution is obtained by integrating
between \vlsr~=~18 and 35~\kms. The lowest contour level and
contour spacing is 4~K~\kms.}
\end{figure}
 
\clearpage
 
\begin{figure}
\vskip -3cm
\figurenum{4}
\epsscale{0.9}
\plotone{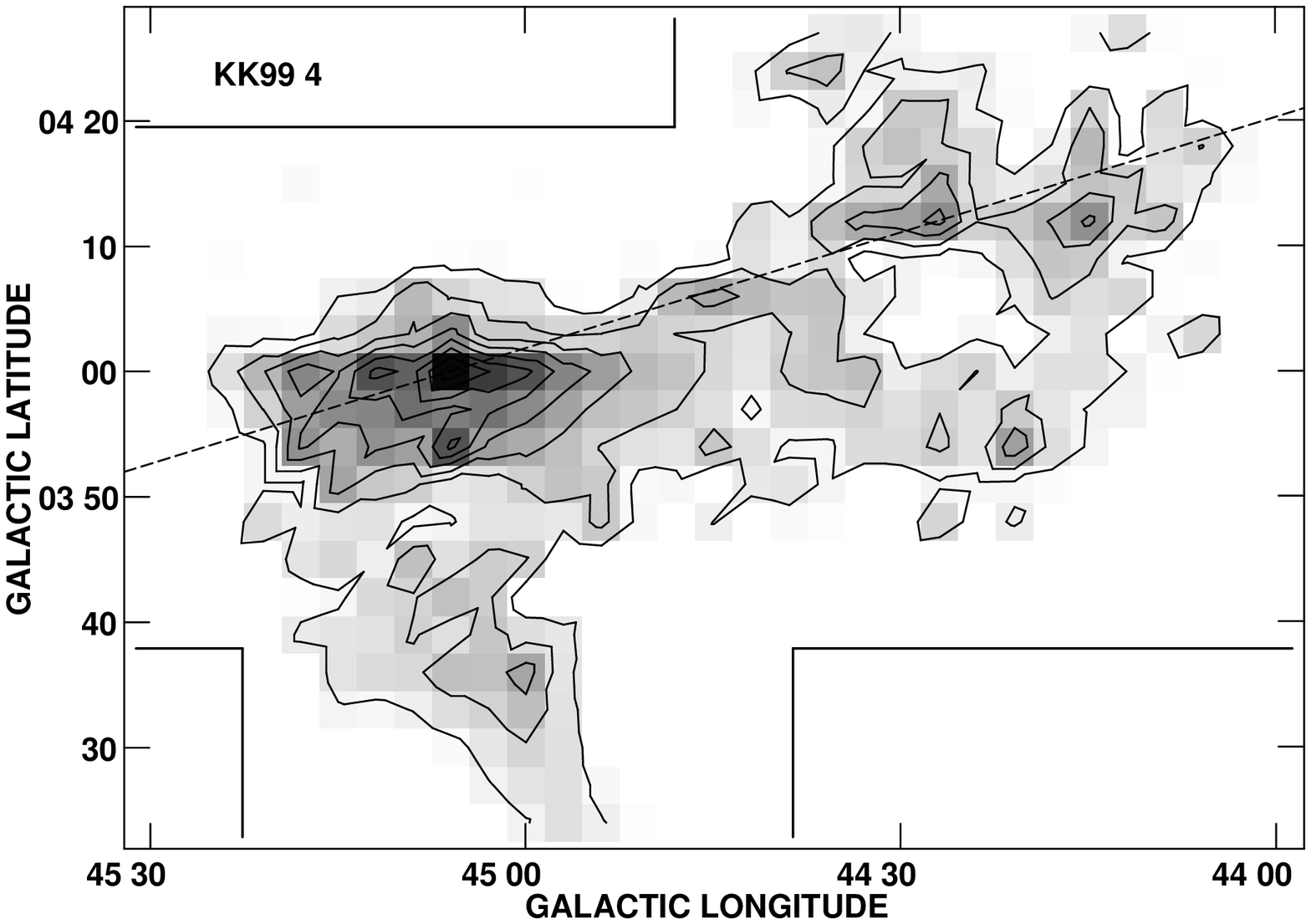}
\vskip -1cm
\caption[figure4b.ps]{
(b) \mctwo. The distribution is obtained by
integrating over the velocity range \vlsr=14$-$25~\kms. The grey scale
levels range from 1 to 18~K~\kms. The lowest
contour level and contour spacing is 2~K~\kms.}
\end{figure}
 
\clearpage
 
\begin{figure}
\vskip -7cm
\figurenum{5}
\epsscale{0.9}
\plotone{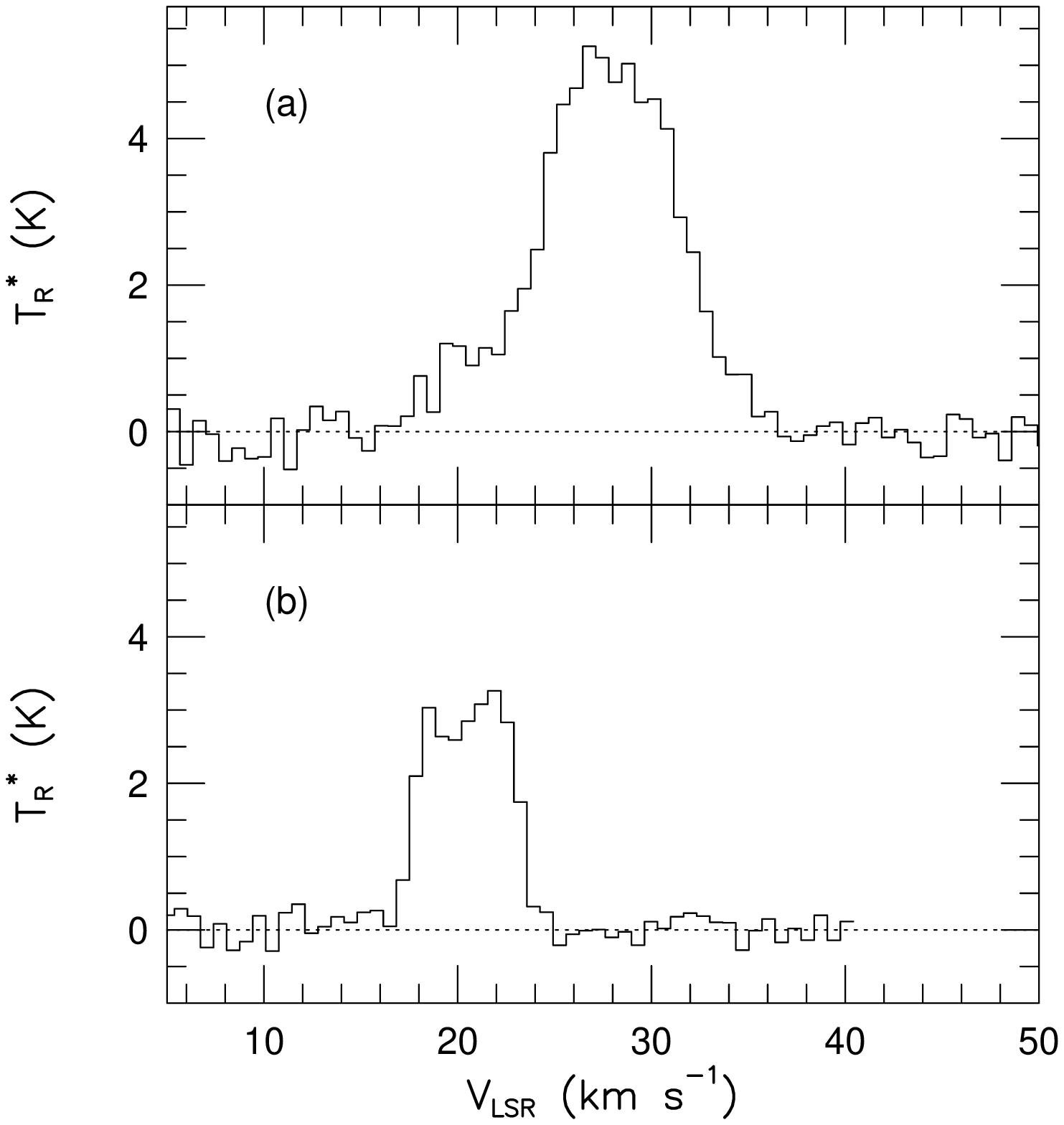}
\vskip -2cm
\figcaption[figure5.ps]{
Peak emission CO line profiles (a) for \mcone\ and (b) for \mctwo.
The line parameters are presented in Table 4.}
\end{figure}
 
\clearpage
 
\begin{figure}
\vskip -3cm
\figurenum{6}
\epsscale{0.8}
\plotone{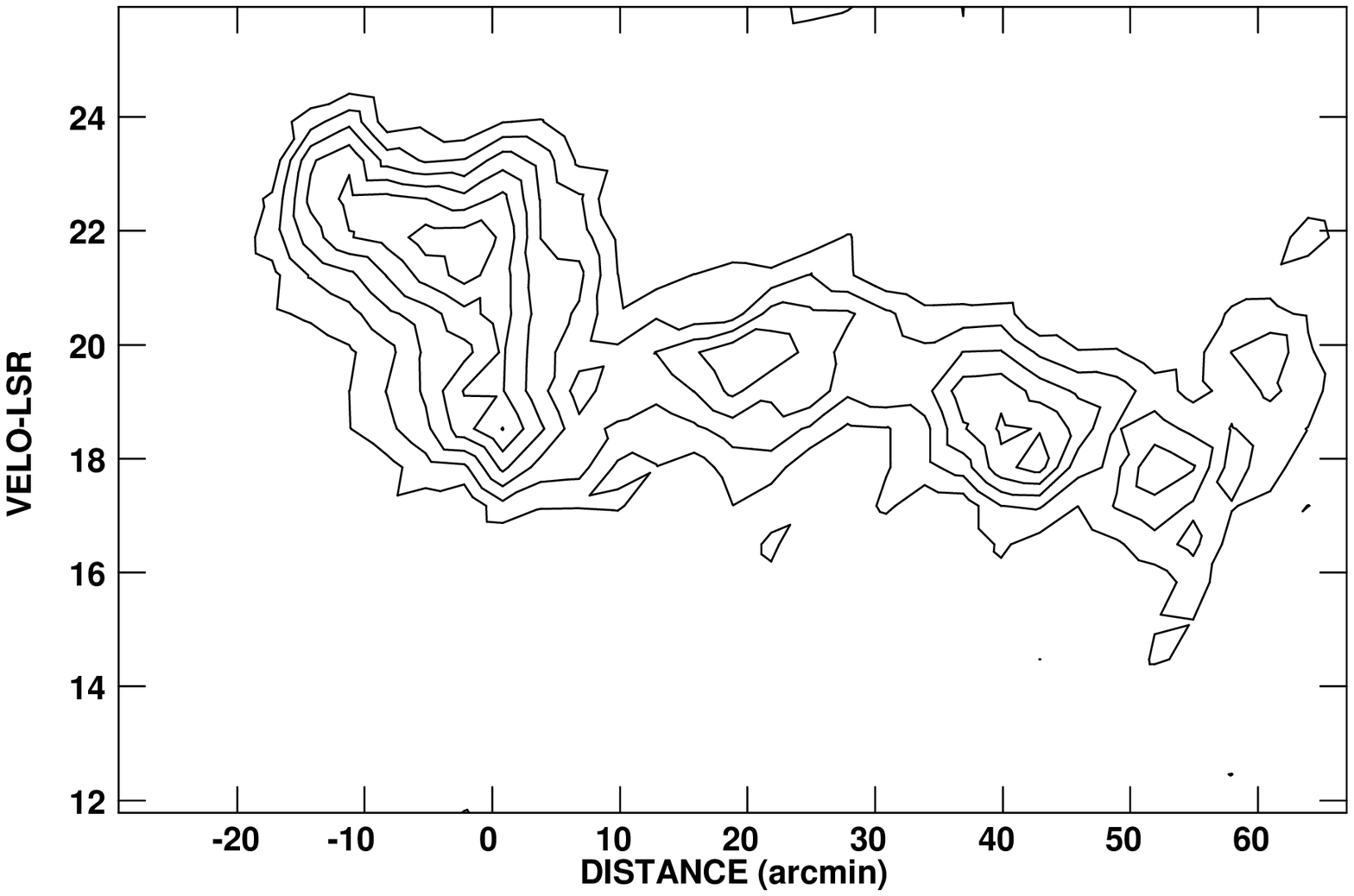}
\caption[figure6.ps]{
Position-velocity diagram of \mctwo\ along the dashed line
in Figure 4(b). The $x$-axis is the distance (positive toward the
northwest) along the line from \lb=(45.\degrs10, 4.\degrs00).
Contour levels are 0.5, 1.0, 1.5, 2.0, 2.5, and 3.0~K}
\end{figure}
 
\clearpage
 
\begin{figure}
\vskip -2cm
\figurenum{7}
\epsscale{0.8}
\vskip -1cm
\plotone{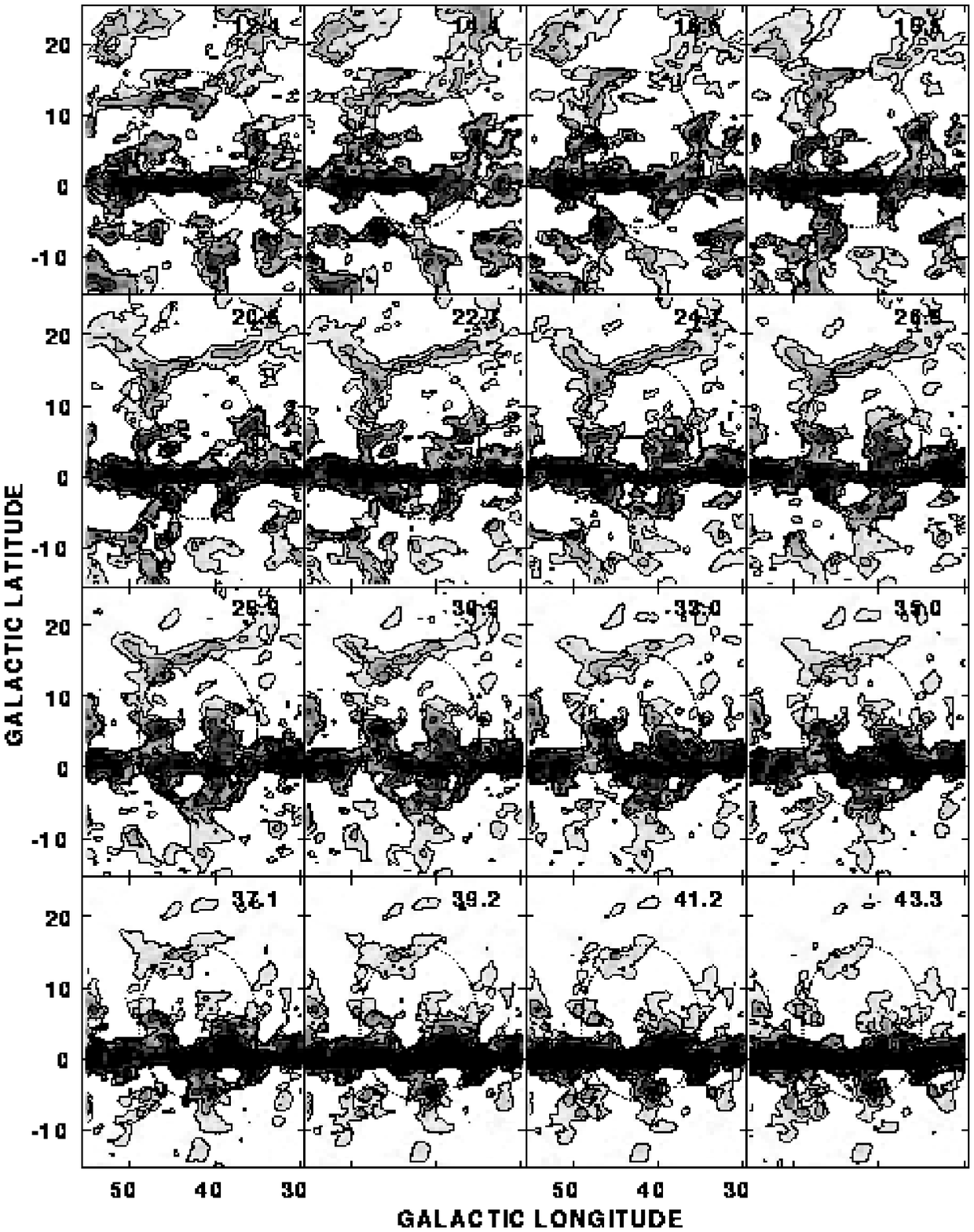}
\caption[figure7.ps]{
Median-filtered HI velocity channel maps,
taken from the Leiden-Dwingeloo survey data.
Each map exhibits the HI intensity distribution in a region of
30\degr$\le l \le$55\degr\ and $-$15\degr$\le b \le$25\degr.
The central velocity is
indicated on the top right corner in each map.
The dotted ellipse centered on \lb=(42\degr, 5\degr)
represents the supershell at
\vlsr=18.5~\kms. The map intensity range
is \ta=0$-$25~K and contours are 1, 5, and 15~K.\\
(A high-quality image file lies in /pub/kimkt at the anonymous ftp site of
star.snu.ac.kr.) }
\end{figure}
 
\clearpage
 
\begin{figure}
\vskip -2cm
\figurenum{8}
\epsscale{0.8}
\plotone{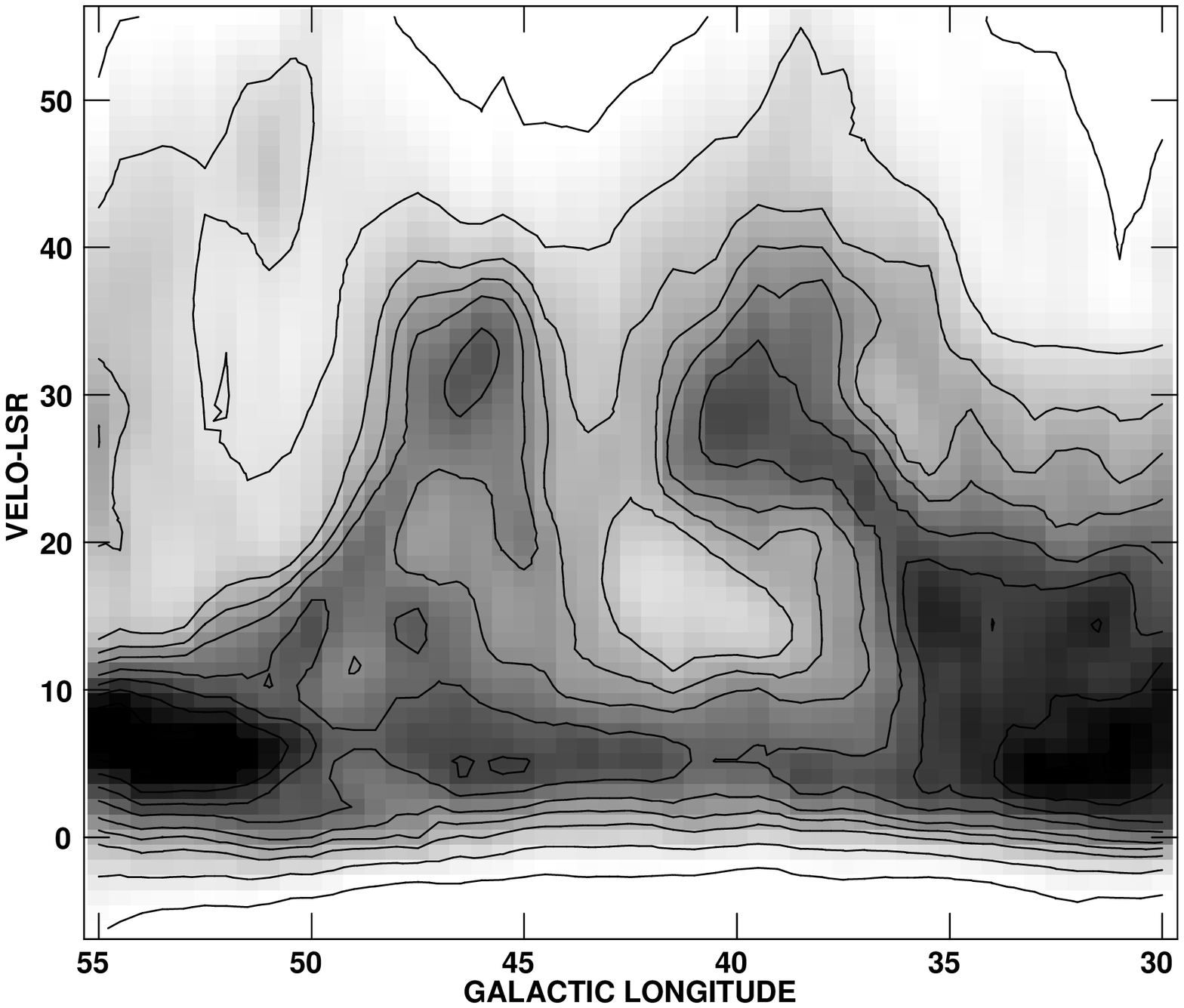}
\vskip -1cm
\caption[figure8.ps]{
Leiden-Dwingeloo HI position-velocity diagram
averaged over $b$=3\degr$-$7\degr.
We can see a hole surrounded by a dense
shell. The hole is centered on ($l$,\vlsr)$\sim$(42\degr, 18~\kms). Two
protuberances correspond to galactic worms \gw\ and GW 39.7+5.7, which
form the vertical walls of the supershell. Contour
levels are 10, 20, 32, 40, 48, 60 ,70, 80, and 90\% of peak intensity,
72~K.}
\end{figure}
 
\clearpage
 
\begin{figure}
\vskip +2cm
\figurenum{9}
\epsscale{2.0}
\plottwo{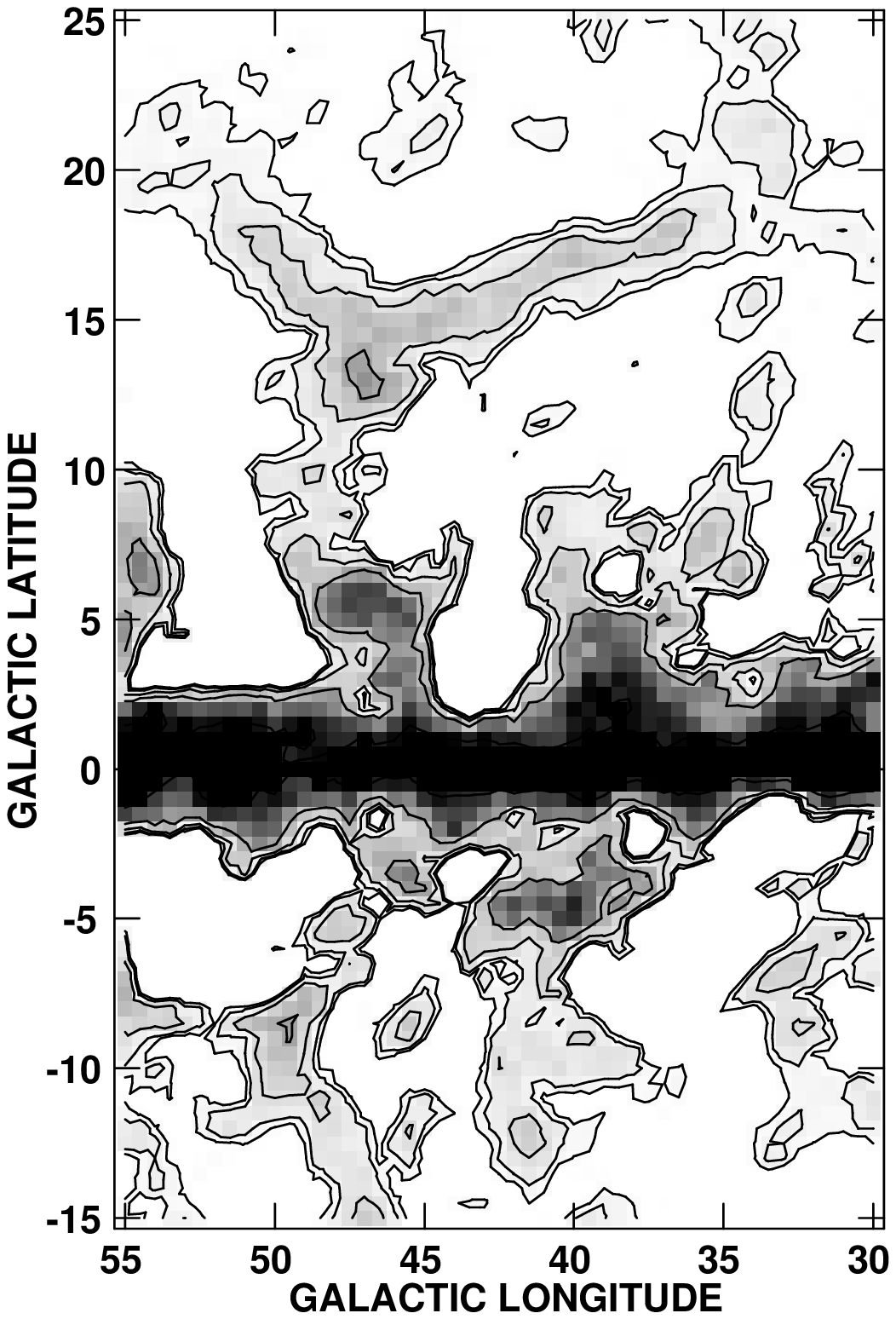}{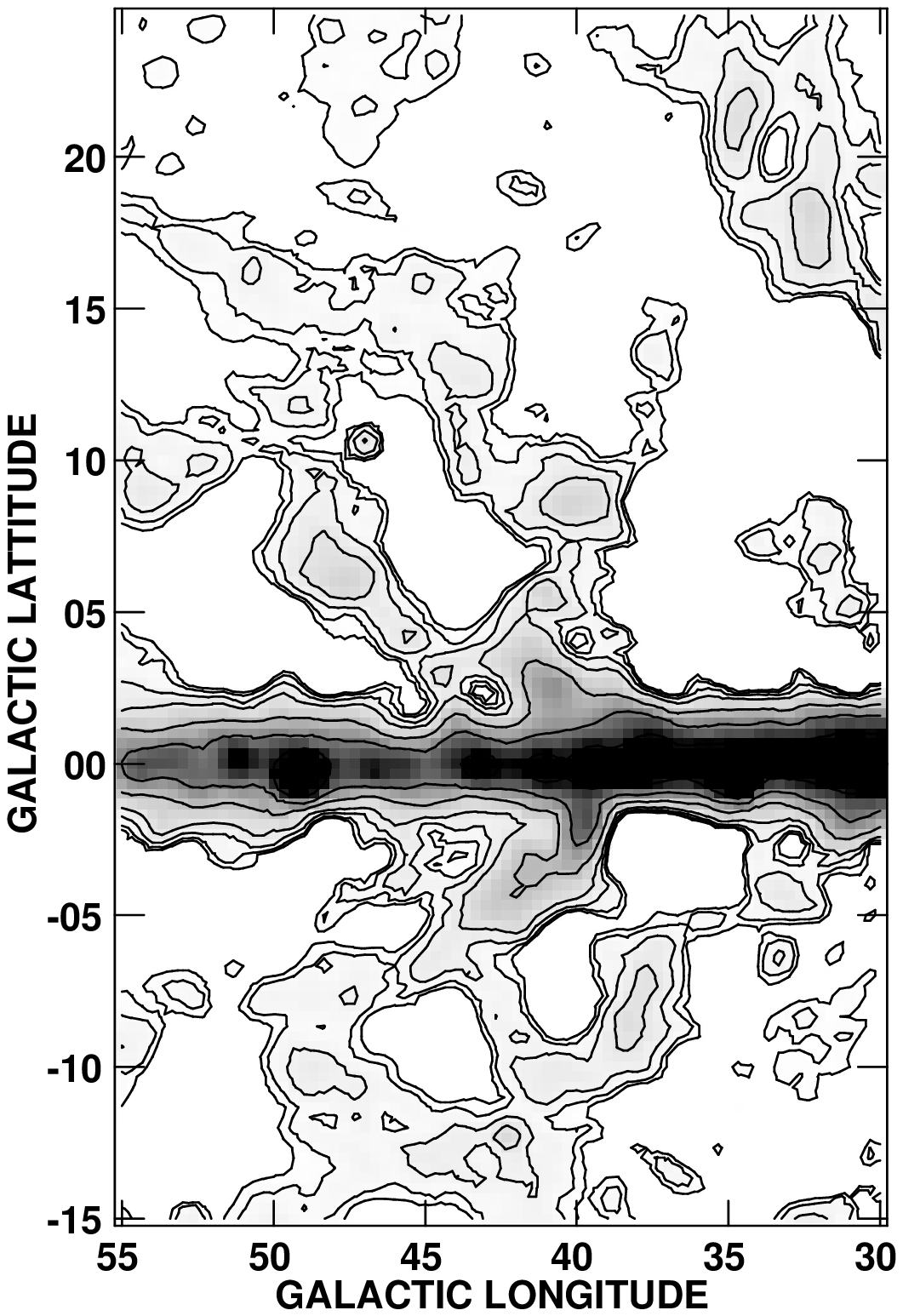}
\vskip -1cm
\caption[figure9.ps]{
(a) Median filtered HI integrated intensity map with a resolution
of 0.\degrs5. Contour levels are 10, 30, 80, and
200~K~\kms\ and grey scale intensity ranges from 0$-$600~K~\kms.
A shell-like structure is seen, centered on \lb$\simeq$(42\degr,
5\degr).
(b) Median filtered 408~MHz continuum map at 0.\degrs85 resolution.
Contour levels are 0.5, 2, 5, 10, 25, 50, 100~K
and grey scale intensity ranges from 0$-$100~K.
The HI shell-like structure has a radio counterpart, which appears as
an almost complete shell.}
\end{figure}
 
\clearpage
 
\begin{figure}
\vskip -5cm
\figurenum{10}
\epsscale{0.9}
\plotone{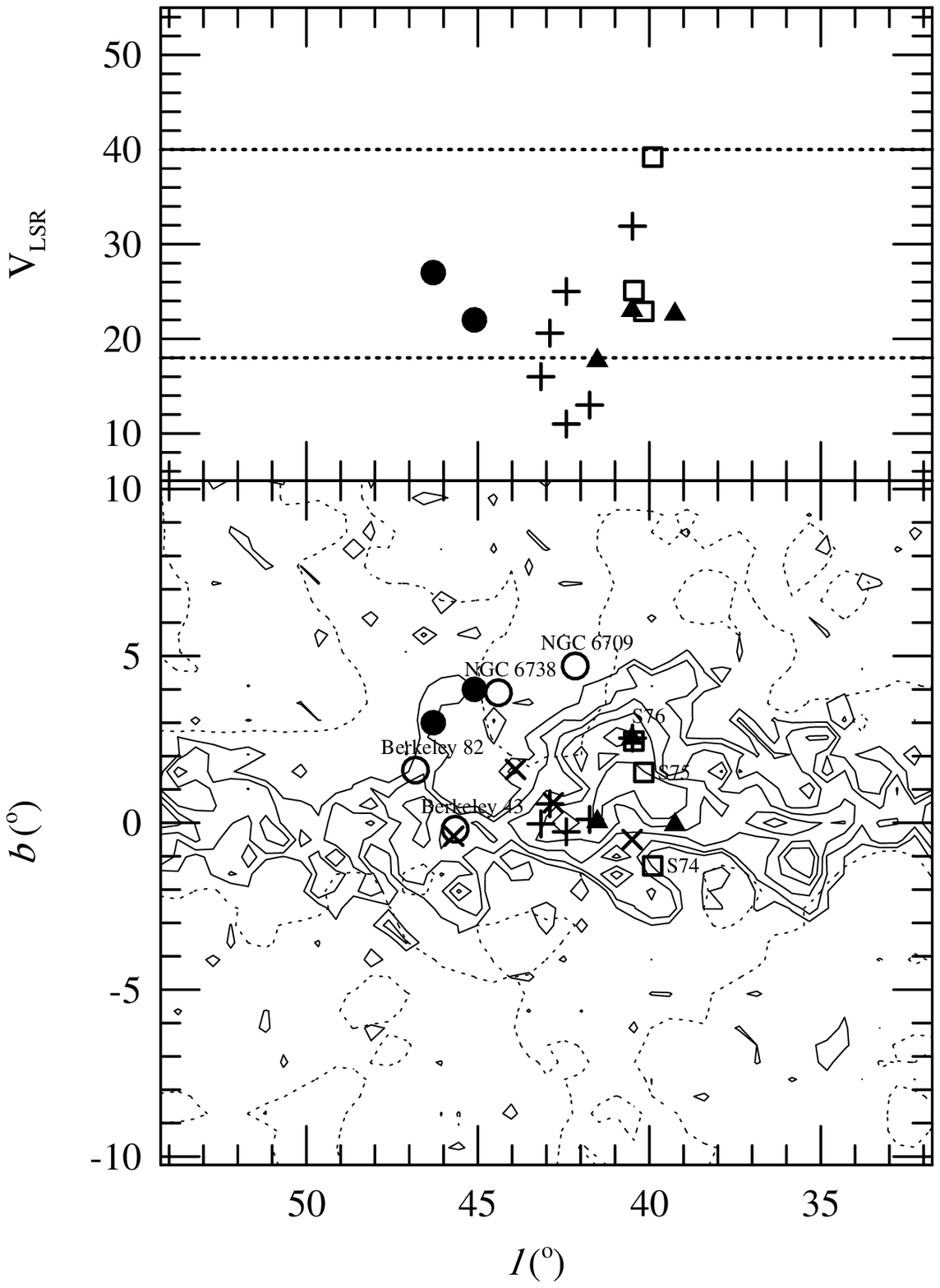}
\vskip -1cm
\caption[figure10.ps]{
{\it Lower panel}: Integrated CO line intensity
at 0.\degrs5 resolution,
obtained by integrating between velocities 18 and 40~\kms.
Contour levels are 2.3, 5, 10, 15, and 20~K~\kms. The
distribution of HI column density is indicated by light dotted lines.
The positions of various objects are also marked. Open circles,
open squares, filled triangles, crosses, and ``$\times$'' symbols
represent open clusters, optical HII regions, radio HII
regions, massive star-forming regions, and SNRs, respectively.
Filled circles indicate \mcone\ and \mctwo, observed in this work.
{\it Upper panel}: The distribution of various objects in the
($l$, \vlsr) plane. Two dotted lines represent the LSR velocity range
of the HI supershell.}
\end{figure}
 
\clearpage

\end{document}